\documentclass{svproc}
\usepackage{amssymb} 
\usepackage{amsmath}       
\usepackage{graphicx}
\usepackage{xcolor}
\usepackage{multicol}     
\usepackage{multirow}   
\usepackage{url}
\usepackage{makecell}

\newcommand{\var}[3]{#1_{#2}(#3)}
\newcommand{\tx}{t}
\newcommand{\X}[1]{\var{X}{#1}{\tx}}

\newcommand{\hX}[1]{\var{\hat{X}}{#1}{\tx}}

\usepackage{algorithm}
\usepackage{algpseudocode}

\usepackage{algorithm}
\usepackage{subcaption}

\begin{document}
\mainmatter              
\title{Conformal based uncertainty bands for predictions in functional ordinary kriging}
\titlerunning{Conformal prediction for functional Ordinary kriging}  
%
\author{Anna De Magistris\inst{1} \and Andrea Diana\inst{1} \and Elvira Romano\inst{1}}
\authorrunning{Anna De Magistris et al.} 
%
\tocauthor{Andrea Diana, Elvira Romano}
\institute{University of Campania Luigi Vanvitelli, Caserta, Italy,\\
\email{anna.demagistris@unicampania.it} \\
\email{andrea.diana@unicampania.it} \\
\email{elvira.romano@unicampania.it}
}

\maketitle              

\begin{abstract}

Functional Ordinary Kriging is the most widely used method to predict a curve at a given spatial point. However, uncertainty remains an open issue. In this article a distribution-free prediction method based on two different modulation functions and two conformity scores is proposed. Through simulations and  benchmark  data analyses, we demonstrate the advantages of our approach when compared to standard methods.
\keywords{conformal prediction, 
 functional ordinary kriging, spatial data.}
\end{abstract}

    \section{Introduction}
        \label{sec:1}
In modern scientific research, the collection of spatially and temporally evolving data has become more common across a wide range of disciplines, including ecology, medicine, biology, geology, and economics. This data, known as spatial functional data \cite{Mateu:17}, consists of functions observed at different spatial locations, capturing intricate variations across both space and time. Spatial Functional Data analysis \cite{Delicado:10} extends classical spatial statistical methods to functional objects \cite{Ramsay:FDA}. One important aspect of spatial functional data analysis is predicting the behaviour of a variable in a location where there are no observations. This challenge has generated different method because Spatial Functional Data has unique characteristics. There are two mainly class of method. The first class analyses the problem taking into account the spatial and temporal autocorrelation of the data, the most famous method are Markov Random Field (MRF) or Spatial AutoRegressive model (SAR) \cite{Cressie:15}. The second class focuses the study of the problem on spatial differences and changing patterns over time, like as Geographically Weighted Functional Regression (GWFR) and  Heteroscedastic Geographically Weighted Functional Regression (HGWFR) (\cite{Diana:23b}, \cite{Romano:20}, \cite{Yamanishi:2003}) or the Ordinary and Universal Kriging (\cite{Delicado:07}, \cite{Menafoglio:13}), where limited ground truth. 

This paper focuses on Kriging methods and explores how uncertainty can be efficiently computed in this context. Kriging involves using information from nearby objects to make predictions at new spatial locations. The predictor's contribution from each function depends on the strength of spatial correlation.
Functional Ordinary Kriging proposed by \cite{Delicado:07} assumes a constant mean function through space and predicts values at unobserved positions based on spatial correlations. Functional Universal Kriging proposed by \cite{Menafoglio:13}, allows spatially varying mean functions and incorporates external covariates into the kriging model to improve prediction accuracy by leveraging additional spatial domain information, such as environmental variables or topographic features. In both cases, the trace-variogram plays a central role in the kriging of spatially correlated functional data (\cite{Mateu:21}, \cite{Mateu:17}). For example, the coefficients of linear combinations of observed functions that define a linear kriging estimate at a new location are determined using a trace-variogram estimate \cite{Giraldo:11}. A recent work \cite{Guo:2022} highlights how the decomposition of trace-variogram into amplitude and phase components allows for more accurate spatial clustering and prediction by accounting for the distinct spatial correlations between amplitude and phase variations, ultimately improving the analysis of misaligned functional data and enabling more robust spatial interpolations.


Despite the advances in kriging techniques for predicting spatial functional data, a significant challenge remains in quantifying the uncertainty of these predictions. Standard bootstrap and smoothed bootstrap methods have been used to obtain confidence intervals for location estimators \cite{Cuevas:06}. Resampling techniques such as bootstrap or permutation methods are typically employed to estimate the uncertainty in predicted curves, allowing for the derivation of confidence bands for functional predictions (\cite{Franco-Villoria:17}, \cite{Ignaccolo:14}). These approaches address the inherent uncertainty in data and prediction processes. While other studies within the FDA framework have provided new insights into this theoretical issue, they have not yet been used in kriging. A first class of methods focuses mainly on the use of parametric bootstrap techniques (\cite{Cao:2012}, \cite{Degras:2011}), while a second class applies reduction of dimensional techniques, useful in dealing with the infinite complexity of the problem (\cite{Antoniadis:2016},  \cite{Hyndman:2007}). 
Both of these classes have significant limitations. They are based on distribution assumptions that are difficult to verify and rely on asymptotic results. In addition, the bootstrap approach is computationally costly. However, ongoing research aims to develop alternative methods that more directly tackle uncertainty in the functional context without strong distributional assumptions. One of such method is Conformal Prediction (CP) (\cite{Diana:23b}, \cite{Diquigiovanni:22}). In this paper we propose a new procedure  which uses Conformal Prediction to define  regions of uncertainty for predictions made through Functional Ordinary Kriging techniques.

The remainder paper is organised as follows. The Kriging methodologies are introduced in Section \ref{sec:2}. The main contribution of the work is presented in Section \ref{sec:3} and discussed  by several simulated and real case studies in Section \ref{sec:4} and Section \ref{sec:5}.


\section{Functional kriging}\label{sec:2} 

Let $\{ X_{s}(t) : s = (u, v) \in \Omega \subseteq \mathbb{R}^{2} \}$ be a geostatistical functional stochastic process  whose functions $X_{s_{i}}\left(t\right)$ are  random functions located in site $s_i$ and for each $s_i=(u_i, v_i)$, where $u_i$ represents latitude and $v_i$ is longitude. Each function is defined on $T=[a,b]\subseteq \mathbb{R}$ and it is assumed to belong to a Hilbert space $ L^2(T) = \{ f: T \rightarrow \mathbb{R} \: \: | \: \: \int_T f(t)^2 dt < \infty\} $ with the inner product $\langle X_{s_{i}}(t),X_{s_{j}}(t) \rangle=\int_T X_{s_{i}}(t)X_{s_{j}}(t) dt$ and norm $ ||X_{s}(t)||_{2} = {\left( \int |X_{s}(t)|^2 dt \right)}^{\frac{1}{2}}$ \cite{Ramsay:FDA}. For a  fixed site $s_{i}$, it is assumed that the observed functions can be expressed according to the model:
 \begin{equation}\label{forma}
	X_{s_{i}}(t)=\mu_{s_{i}}(t)+\epsilon_{s_{i}}(t),\ \ i=1,\ldots,n
	\end{equation}
 
where deterministic component $\mu_{s_{i}}\left(t\right)$ describes spatial mean variation and a stationary stochastic component $\epsilon_{s_{i}}(t)$ is supposed to be a zero-mean. Two types of kriging are distinguished in relation to the stationary of the process: Ordinary Kriging (OK) and Universal Kriging (UK). 

In according to \cite{Giraldo:11} and \cite{Goulard} the definition of the Kriging predictor for functional data to estimate the variable $X_{s_{0}}(t)$ at location $s_0 \in \Omega$ is the best linear unbiased predictor (BLUP): 

\begin{equation}
		X_{s_{0}}^*(t)=\sum_{i=1}^{n}\lambda_i^*X_{s_{i}}(t)
	\end{equation}

whose weights $\lambda_1^*,\dots,\lambda_n^*$ minimize the global variance of the prediction error under the unbiasedness constraint:

\begin{equation} \label{solution}
(\lambda_1^*,\dots,\lambda_n^*) = \underset{\lambda_1,\dots,\lambda_n \in \mathbb{R}}{\operatorname{argmin}}  \mathbb{V}(X_{s_{0}}^{*}(t) - X_{s_{0}}(t)) \: \: \: t.c. \: \: \mathbb{E}[X_{s_{0}}^{*}(t) - X_{s_{0}}(t)] = 0
\end{equation}

To solve equation \eqref{solution}, we estimate the semivariogram model, denoted by \(\gamma(h)\), where \(h\) represents the spatial distance between two sites \(s_i\) and \(s_j\) in the domain \(\Omega\). The semivariogram is a measure of the variation in dissimilarity between observations as a function of the spatial distance between them. 

The semivariogram \(\gamma(h)\) is defined as half of the variance of the difference between two observations \(X_{s_{i}}\) and \(X_{s_{j}}\), or equivalently, as the squared expected difference between them:
\[ \gamma(h,t) = \frac{1}{2} \mathbb{V}(X_{s_{i}}(t) - X_{s_{j}}(t)) = \mathbb{E}[(X_{s_{i}}(t) - X_{s_{j}}(t))]^2.\]
Using Fubini’s theorem, this relationship can be expressed as:

\begin{equation} \label{eq4}
    \gamma(h) = \frac{1}{2}  \mathbb{E}[ \: ||X_{s_{i}}(t) - X_{s_{j}}(t)||^2]
\end{equation}

for $s_i, s_j \in \Omega$ and $h = ||s_i-s_j||$.

Overall, it provides important information about the spatial correlation structure of the data. After discussing the general formulation of the kriging prediction problem for functional data and the importance of semivariogram model estimation, we analyze the particular case of Ordinary Kriging.

\subsection{Functional ordinary kriging}

In the case of Ordinary kriging, the stochastic process is second-order stationary and isotropic i.e.:

\begin{itemize}
		\item[•] $\mathbb{E}[X_{s}(t)]=\mu(t), \ \forall s \in \Omega;$
        \item[•] $\mathbb{V}[X_{s}(t)] = \sigma^{2}, \forall
        s \in \Omega; $
		\item[•] $Cov(X_{s_{i}}(t), X_{s_{j}}(t))=\mathbb{E}[\langle X_{s_{i}}(t),X_{s_{j}}(t) \rangle]=C(h),\ \ \forall i,j=1,\ldots,n,\ \ h=||s_i-s_j||;$
	\end{itemize}
 which implies that the mean and variance of the process are constant throughout the spatial domain.

The unbiaseness condition $ \mathbb{E}[\epsilon_{s_{0}}(t)] = 0 $ in \eqref{solution} results in  $\sum_{i=1}^{n} \lambda_i = 1 $:
	
\begin{align*}
    \mathbb{E}[X_{s_{0}}^{*}(t) - X_{s_{0}}(t)] &= \mathbb{E} \left[ \sum_{i=1}^{n} \lambda_i^* X_{s_{i}}(t) - X_{s_{0}}(t) \right] =  \\
    &= \sum_{i=1}^{n} \lambda_i^* \mathbb{E}[X_{s_{i}}(t)] - \mathbb{E}[X_{s_{0}}(t)] =  \\
    &= \sum_{i=1}^{n} \lambda_i^* \mu - \mu = \\
    &= \mu \left( \sum_{i=1}^{n} \lambda_i^* - 1 \right) .
\end{align*}

 Using the Lagrange multiplier method the optimization problem \eqref{solution} can be traced back to the determination of the optimal weights $\lambda_1^*,\dots,\lambda_n^*$ that minimize functionality: 

\begin{align} \label{sol}
    \phi = \mathbb{V}(\epsilon_{s_{0}}) + 2 L \left(\sum_{i=1}^{n} \lambda_i - 1 \right) = \sum_{i=i}^{n}\sum_{j=1}^{n} \lambda_i \lambda_j C(h_{i,j}) + \\ 
    \notag
    + C(0) - 2\sum_{i=i}^{n} \lambda_i C(h_{i,0}) + 2 L \left(\sum_{i=1}^{n} \lambda_i - 1\right)
\end{align}

and taking into account that $ \gamma(h) = C(0)-C(h)$ the minimum of the functionality \eqref{sol} can be identified by imposing that the partial derivatives are equal to zero, obtaining the following system:

\begin{equation}\label{sistem}
\begin{pmatrix}
\gamma(0) & \gamma(h_{1,2}) & \dots & \gamma(h_{1,n}) & 1 \\
\gamma(h_{2,1}) & \gamma(0) & \dots & \gamma(h_{2,n})& 1 \\
\vdots & \vdots & \ddots & \vdots & \vdots \\
\gamma(h_{n,1}) & \gamma(h_{n,2}) & \dots & \gamma(0) & 1 \\
1 & 1 & \dots & 1 & 0 \\
\end{pmatrix}
\begin{pmatrix}
\lambda_1 \\
\lambda_2 \\
\vdots \\
\lambda_n \\
m \\
\end{pmatrix}
=
\begin{pmatrix}
\gamma(h_{0,1}) \\
\gamma(h_{0,2}) \\
\vdots \\
\gamma(h_{0,n}) \\
1 \\
\end{pmatrix}
\end{equation}

where $\gamma(h_{i,j}) = \frac{1}{2}  \mathbb{E}[ \: ||X_{s_{i}}(t) - X_{s_{j}}(t)||^2] $
for $s_i, s_j \in \Omega$.
The system \eqref{sistem} admits a unique solution if and only if the matrix 
   $ \begin{pmatrix}
        \gamma(h_{i,j}) & 1\\
        1 &  0\\
    \end{pmatrix} $
is non-singular. This implies that the semivariogram must be defined as positive, so that the coefficient matrix is invertible. The semivariogram must be consistent with the spatial structure of the data and must satisfy the properties of positivity and symmetry.
Furthermore, the system solution provides the optimal weights $\lambda_1^*,\dots,\lambda_n^*$ and the Lagrange multiplier $L$ that minimise the global variance of the prediction error while ensuring fairness of the predictor. 

One important challenge related to these prediction errors is to assess the uncertainty. 
 Techniques 
 such as resampling methods (\cite{Antoniadis:2016}, \cite{Cao:2012}, \cite{Degras:2011},  \cite{Hyndman:2007}) have been introduced  to estimate the uncertainty in predicted curves, allowing for confidence bands for functional predictions. However these are computationally intensive, especially for large datasets, due to the need for multiple resampling iterations. In addition, they may lead to biased estimates if the sample size is small or not representative of the population.
A direct mechanism for constructing confidence intervals that are guaranteed to have the desired coverage probability is the Conformal Prediction.
Adaptable to various data structures, including functional and spatial data,
CP is a framework that generates prediction intervals. It ensures that the intervals contain the true value with a specified confidence level, based on the distribution of errors from a training set. The main focus of this work is to present this innovative forecasting method within Functional Ordinary Kriging. Integrating conformal prediction with kriging provides a robust framework for generating reliable prediction intervals in spatial statistics. 
 
 \section{Conformal prediction for functional Ordinary Kriging}
\label{sec:3}

Conformal Prediction aims to create predictive regions that are expected to contain the true outcome with a specified level of confidence (denoted as $\alpha$). These regions, denoted as $C(x)$ for $x \in \mathbb{R}^n$, are designed so that the probability of the true outcome falling within the region is at least $1 - \alpha$. Calibration is important step in Conformal Prediction, ensuring that methods are properly calibrated if they maintain a probability of at least $1 - \alpha$ for the true outcome falling within the predicted region. This statistical integrity holds regardless of whether the underlying predictor comes from statistical, machine learning, or deep learning methods, providing reliable uncertainty assessments. For a more detailed discussion, refer to (\cite{Mao1}, \cite{Vok:2005}).

In functional data analysis, Conformal Prediction emerges as a robust machine learning framework for uncertainty quantification. It focuses on generating prediction regions, commonly referred to as prediction bands, that maintain statistical validity for any underlying point predictor. A first approach for spatial functional data is proposed in \cite{Diana:23b} for regression models.

In this section we define the theoretical basis of conformal prediction for Functional Ordinary Kriking based on two different modulation functions and two conformity scores.

Let $(\Omega, \mathcal{F}, \mathbb{P})$ be a probability space, where $\Omega$ represents the sample space, $\mathcal{F}$ is the sigma-algebra (a collection of subsets of $\Omega$ that defines measurable events), and $\mathbb{P}$ is the probability measure that maps events in $\mathcal{F}$ to probabilities in $[0, 1]$. 

Let $\{ X_{s}(t) : s = (u, v) \in \Omega \subseteq \mathbb{R}^{2} \}$ be a geostatistical functional stochastic process, to facilitate notation we indicate the data with $\boldsymbol{z}_{s_i}=(s_i, \X{s_i})$ with  $ t \in T$ and each pair representing one sample. 

To evaluate the uncertainty of a predicted curve $X^*_{s_{0}}(t)$ from a new site $s_0$ without making any assumptions about the distribution of data we use Conformal Prediction. A valid prediction set for $\boldsymbol{z}_{s_{0}} = (s_{0}, X^*_{s_{0}}(t))$, which is assumed to be spatial exchangeability as $\boldsymbol{z}_{s_{1}}, \ldots, \boldsymbol{z}_{s_{n}}$, is the set $ C \subset\Omega\times L^2(T)  $ constructed based on $\boldsymbol{z}_{s_{1}}, \ldots, \boldsymbol{z}_{s_{n}}$ such that

\begin{equation*}
      \mathbb{P}(X_{s_{0}}(t) \in C(s_{0})) \geq 1- \alpha
\end{equation*}
for any significance level $\alpha \in (0, 1)$.
CP operates under the assumption of spatial exchangeability that  can be formally defined as:
\begin{definition}[Spatial exchangeability] 
An spatial exchangeable sequence of random functions is a finite or infinite sequence 
\[\{(s_1, X_{s_{1}}(t)), (s_2,X_{s_{2}}(t)), \dots, \linebreak (s_n,X_{s_{n}}(t))\}\]
of random functions such that for any finite permutation \(\sigma\) of the site \(\{1, 2,\ldots,n\}\), the joint probability distribution of the permuted sequence 
\[\{(s_{1},X_{s_{\sigma(1)}}(t)), (s_2,X_{s_{\sigma(2)}}(t)), \linebreak \dots , (s_n,X_{s_{\sigma(n)}}(t))\}\] is the same as the joint probability distribution of the original sequence.
\end{definition}

Split Conformal approach is used to construct prediction bands. Given data $\boldsymbol{z}_{s_{1}}, \ldots, \boldsymbol{z}_{s_{n}}$, randomly divide the set $\{1, \ldots, n\}$ into two subsets $I_1$ and $I_2$. Define the training set as $Z_{TRAIN}:=\{\boldsymbol{z}_{s_{h}} : h \in I_1\}$ and the calibration set as $Z_{TEST}:=\{\boldsymbol{z}_{s_{k}} : k \in I_2\}$, where $|I_1| = m$, $|I_2| = l$, and $m, l \in \mathbb{N}_{>0}$ such that $n = m + l$. We then define a non-conformity measure as any measurable function $\mathcal{D}(\{\boldsymbol{z}_{s_{h}} : h \in I_1\}, \boldsymbol{z}_{s_{}})$ that takes values in $\overline{\mathbb{R}}$, the set of affinely extended real numbers. The Split Conformal prediction set for $\X{s_0}$ is then defined as
\[
C(s_{0}) := \{\X{s} \in L^2(T) : \delta_{\X{s}} > \alpha\}
\]

with 
\[
\delta_{\X{0}} := \frac{1}{l+1} \left| \{k \in I_2 \cup \{0\} : R_k \geq R_{0}, \} \right|
\]

and nonconformity scores $R_k := \mathcal{D}(\{z_h : h \in I_1\}, z_k)$ for $ k \in I_2$, $ R_{0} := \mathcal{D}(\{z_h : h \in I_1\}, (s_0,X^*_{s_{0}}(t)))$.
According to \cite{Diquigiovanni:22} we define the prediction band as follows:

 \[
			C(s_{0})=\{ X_{s}(t) \in {L}^2(T): X_s(t) \in [X^*_{s_{0}}(t) - \rho^{\mathcal{S}}\mathcal{S}(\tx), X^*_{s_{0}}(t) + \rho^{\mathcal{S}}\mathcal{S}(\tx)],\; \forall t \in T\} \]

where $S(t)$ is a modulation function, $\rho^s$ is the ray of prediction band and $X^*_{s_{0}}(t)$ is the mean of the estimated model and the centre of prediction band. In particular $\rho^s$ is the value of $(1 - \alpha)$ quantile distribution of non-conformity measure. 

The modulation function adjusts the width of the prediction band for the functional variable $X(t)$, enabling adaptive modelling to local variations and specificity in the functional data. This construction of the prediction band, guided by $\mathcal{S}(t)$, contributes to targeted uncertainty quantification, providing more informative and adaptable predictions in complex analytical contexts. The modulation function $\mathcal{S}(t)$ is crucial for adapting the width of the prediction band to local variations in the functional data. This feature can be designed based on the data structure and specifics of the problem, for example to reflect seasonal variations, time cycles, or local trends in the functional data. An appropriate choice of $S(t)$ can improve model fit and prediction accuracy. The modulation functions used are:
        \begin{align}
            \mathcal{S}_{sup}(\tx):=sup_{\hat{X}_{s_{0k}}\in \Tilde{X}}| X^*_{s_{0}}(t) - \hX{s_{0j}}|; \label{mod_sup}  \\
            \mathcal{S}_{sqrt}(\tx):=\sqrt{\frac{\sum_{\hat{X}_{s_{0k}}\in \Tilde{X}}(X^*_{s_{0}}(t) - \hX{s_{0j}})^2}{|\Tilde{X}|}}, \label{mod_sqrt}
        \end{align}
where $\hat{X}_{{s_{0j}}} $ is the prediction with  Kriging on $Z_{TEST}$. This two function are used because $\mathcal{S}_{sup}(\tx)$, defined in (\ref{mod_sup}), measure locally the variability around the \(X^{*}_{s_{0}}(\tx)\); on the contrary, $\mathcal{S}_{sqrt}(\tx)$, defined in (\ref{mod_sqrt}), measure globally the variability around the \(X^{*}_{s_{0}}(\tx)\).
Instead, the non conformity measures used are:
   \begin{align}
        \mathcal{D}_{sup}\left(\frac{\hat{X}_{s_{0j}}(\tx)}{\mathcal{S}(\tx)}; \frac{X^*_{s_{0}}(\tx)}{\mathcal{S}(\tx)}\right):=sup_{\tx\in T}\left| \frac{X^*_{s_{0}}(t) - \hX{s_{0j}}}{\mathcal{S}(\tx)} \right|\label{nnc_sup}
        \\
        \mathcal{D}_{sqrt}\left(\frac{\hat{X}_{s_{0j}}(\tx)}{\mathcal{S}(\tx)}; \frac{X^*_{s_{0}}(\tx)}{\mathcal{S}(\tx)}\right):= \sqrt{  \int{ \frac{ (X^*_{s_{0}}(t) - \hX{s_{0j}})^2 dt}{\mathcal{S}(\tx)}} }.\label{nnc_sqrt}
   \end{align}
This two function are used because $\mathcal{D}_{sup}$, defined in (\ref{nnc_sup}), measure locally the distance between predicted values and \(X^{*}_{s_{0}}(\tx)\) and is the non conformity measures introduced in the previous section and defined by \cite{Diquigiovanni:22}; on the contrary, $\mathcal{D}_{sqrt}$, defined in (\ref{nnc_sqrt}), measure globally distance between predicted values and \(X^{*}_{s_{0}}(\tx)\).

The modulation functions, such as \( \mathcal{S}_{sup}(\tx) \) and \( \mathcal{S}_{sqrt}(\tx) \), adjust the width of the prediction band based on local variations in the data. Meanwhile, the non-conformity measures, like \( \mathcal{D}_{sup} \) and \( \mathcal{D}_{sqrt} \), quantify the deviation between predicted values around \(X^{*}_{s_{0}}(\tx)\).

The parameter $\rho^{\mathcal{S}}$ represents the radius of the prediction band and is determined as the value of the $(1 - \alpha)$ quantile of the nonconformity measure. This measure of nonconformity reflects the discrepancy between model predictions and observed data, allowing you to calculate a prediction band that takes into account residual variability not explained by the model. The conformal prediction band provides an estimate of the uncertainty associated with predicting the functional curve $X_{s_0}(t)$ at the new site $s_0$. The probability $(1-\alpha)$ associated with the prediction band indicates the probability that the actual functional curve falls within the interval, providing an assessment of confidence in the accuracy of the prediction.

\subsection{Procedure}

The main steps of the Conformal Prediction procedure for functional Ordinary Kriging involve several stages that aim to provide reliable predictions along with quantified uncertainty. We divide the dataset into two subset: the training set ($Z_{TRAIN}$) and the test set ($Z_{TEST}$). This division is important for calibrating the model and creating prediction bands. To do this, we follow the principle that nearby observations are more related than distant ones, as suggested by Waldo Tobler: ``everything is related to everything else, but nearby things are more related than distant ones". So, we identify the $K$ closest sites to our target prediction location $s_0$, forming a proximity set $Prox(s_0)$:
\[Prox(s_{0})=\{\boldsymbol{z}_{s_i} : h_{i,0}<\Delta_K\}\]

where $\Delta_K$ is a proximity threshold. 
To split the dataset to have the same order $\Delta_K$ is set to the median value of spatial distances from \(s_0\).
The training set \(Z_{TRAIN}\) is the proximity set, while the test set \(Z_{TEST}\) contains the remaining observations:
\[Z_{TRAIN}:= Prox(s_{0}), \: \: \: \: Z_{TEST}= Z \smallsetminus Z_{TRAIN}.\]

The idea of conformal prediction is to try all possible curves for the test object to see how well these curves conform to the set of training examples.
We use the training set \(Z_{TRAIN}\) to estimate the trace-variogram with errors. This is an important step for kriging because it helps us understand how the data is spatially correlated. The estimated trace-variogram allows us to create a model for the covariance between observations at different locations. The theoretical trace-variogram can be estimated as:
\[\hat{\gamma}(h) = \frac{1}{|2N(h)|}\sum_{(i,j) \in N(h)} ||X_{s_{i}}(t) - X_{s_{j}}(t) ||^2\] 

where $ N(h) = \{ (i,j) \: \: : ||s_i -s_j||=h \} $ and $ |N(h)|$ indicates the cardinality.
After estimating the trace-variogram, we use data from \(Z_{TRAIN}\) to predict the target variable \(X^{*}_{s_{0}}(t)\) at the prediction location \(s_{0}\). We use ordinary kriging for this, which utilizes the spatial correlation information obtained from the trace-variogram as explained in Section~\ref{sec:2}.
We use the test set \(Z_{TEST}\) to determine the radius and modulation for the prediction band for \(X^{*}_{s_{0}}(t)\). To create the prediction bands, we consider different combinations of modulation functions \eqref{mod_sup}, \eqref{mod_sqrt}, and nonconformity measures \eqref{nnc_sup}, \eqref{nnc_sqrt} as defined in the previous section. The radius of the prediction band is set as the value of the $(1 - \alpha)$ quantile of the nonconformity measure.
We construct the forecast band based on the determined radius and modulation function for \(X^{*}_{s_{0}}\):
\[C(s_{0})=\{ X_{s}(t) \in L^2(T): X^*_{s_{0}}(t)-\rho^{\mathcal{S}}\mathcal{S}(\tx)\le f(\tx) \le X^*_{s_{0}}(t)+\rho^{\mathcal{S}}\mathcal{S}(\tx)\;\forall\tx\in T\}.\]


To implement the conformal prediction method in the context of functional Ordinary kriging, we adopt Algorithm  \ref{algo:1}  below. This algorithm allows you to construct a prediction band for a functional variable, using observed data to fit the model and generate a predictive estimate along with an indication of the associated uncertainty. The algorithm follows a series of steps, including prediction with ordinary kriging, building a dataset for test-based prediction, and calculating the prediction band radius and modulation function.

\begin{algorithm}[H]
			\caption{Conformal Prediction}\label{algo:1}
			\begin{algorithmic}
				\Require $Z_{TRAIN}$ observations, $Z_{TEST}$ observations, $s_0$ prediction position.

				\Ensure  $ \rho^{\mathcal{S}}$ prediction band radius, modulation function $\mathcal{S}$.
    
\State $X^*_{{s_{0}}}(t) \gets $ prediction with Ordinary Kriging on $Z_{TRAIN}$
\State  $ \Tilde{X}(t) \gets$ construction of the set as follows
				\For{$\boldsymbol{z}_{s_j} \in Z_{TEST}$}
				
			 \State $\hat{Z} \gets \{Z_{TRAIN},\boldsymbol{z}_{s_j}\}$
            \State $\hat{X}_{{s_{0j}}}(t) \gets $ prediction with Ordinary Kriging on $\hat{Z}$
				\State $\Tilde{X}_j(t) \gets \hat{X}_{{s_{0j}}}(t) $
				\EndFor
				\State $\mathcal{S}(\tx) \gets $ define the modulation function 
\For{$\Tilde{X}_{s_{0j}}(t) \in \Tilde{X}$ }
            \State $R_j \gets \mathcal{D}\left(\frac{\hat{X}_{s_{0j}}(\tx)}{\mathcal{S}(\tx)}; \frac{X^*_{s_{0}}(\tx)}{\mathcal{S}(\tx)}\right)$ non-conformity scores
            \EndFor
\State  $\rho^{\mathcal{S}} \gets $ $(1-\alpha)$-th percentile of distribution of $R_j$.
				
			\end{algorithmic}
		\end{algorithm}

In the kriging procedure for solving the system \eqref{sistem}, both classical and iterative methods can be used to solve the system of equations needed to calculate the estimates. In our case we used the conjugate gradient method. The conjugate gradient is particularly useful in dealing with ill-conditioning problems generated by the local choice of the train set, i.e. the fact that the available data points may be non-uniformly distributed or have complex spatial correlations. This can cause problems with numerical stability and accuracy in estimates, especially when trying to solve the system of equations to obtain the optimal weights for interpolation. The conjugate gradient offers several advantages in this context. First, it can effectively handle ill-conditioning issues, improving numerical stability and reducing the risk of instability when calculating estimates. Furthermore, being an iterative method, it can be more efficient in terms of computational resources than classical methods, especially when working with large datasets or sparse matrices. The use of the conjugate gradient in the kriging procedure allows the ill-conditioning problems arising from the local choice of the train set to be addressed more effectively, improving the precision and stability of the interpolative estimates.

    \section{Simulated case studies}
        \label{sec:4}

To evaluate the performance of the algorithm, in terms of bandwidth coverage and computation time, we use simulated data.  We  measure the effectiveness of our method using various key metrics. This test help us to identify the advantages and limitations of each approach, offering a comprehensive analysis of performance across different application scenarios.

\subsection{Key metrics}
The performance evaluation of conformal prediction for the Kriging method involves several metrics:

\begin{itemize}
  
    \item[•]   
    The Band width, $Width:=\int_T(I_u(t)-I_l(t))dt$ where $I(t)=[I_u(t),I_l(t)]$ represents the prediction band. It providing an approximate margin of error  equal to $Width/2$;
    \item[•] 
        The Band Score $S_{\alpha}$, defined as:
\[  S_{\alpha}(I(t),X_{s_i}(t))= \int_T A(I(t),X_{s_i}(t)) dt\]
    where
    \begin{equation*}
        A(I(t),X_{s_i}(t))  = (I_u(t)-I_l(t)) + \frac{2}{\alpha}(I_l(t)-X_{s_i}(t))_+ + \frac{2}{\alpha}(X_{s_i}(t)-I_u(t))_+
    \end{equation*}
    and $z_+=z\lor 0$ denotes the “positive part”. A smaller $S_{\alpha}$ is desirable as this rewards both high coverage and narrow intervals. The minimum of $S_{\alpha}$ is equal to the $Width$ index when the data are completely contained within the band; otherwise, $S_{\alpha}$ measures the distance of the data from the band.

      \item[•] 
        The Global Coverage of the $(1-\alpha)100\%$ prediction band $Cov{\alpha_{G}}$, which represents the percentage of curves contained within the prediction band;
    \item[•] 
        The Local coverage of the prediction band $Cov{\alpha_{L}}(t)$, indicating the percentage of curve points within the prediction band. It quantifies how many points satisfy $X_{s_i}(t)\in [I_l(t), I_u(t)]$, defined as:

\begin{equation}					
Cov\alpha_L(t) = \frac{1}{N} \sum_{i=1}^{N} \mathbb{I}\left( X_{s_i}(t)\in [I_l(t), I_u(t)] \right)	
\end{equation}			
	where 
\[\mathbb{I}\left( X_{s_i}(t) \in [I_l(t), I_u(t)] \right) = 
\begin{cases} 
    1 & \text{if } I_l(t) \leq X_{s_i}(t)\leq I_u(t), \\
    0 & \text{otherwise};
\end{cases}\]

    \item[•] Total time (TT), which represents the total time to build the prediction band for each curve in the dataset.
          \item[•] Mono time (MT), which represents the average time to build the forecast band for a single curve in the dataset.
\end{itemize}

 \subsection{Set up algorithm parameters}
We evaluate simulated and real data combining different parameters of the proposed algorithm. In particular, we employ three different proximity thresholds to partition the dataset into training and testing sets:
        
\begin{enumerate}
    \item 
        $\Delta_{25}$, defined as the value of $25-$th percentile of $\{h_{i,0}:i=1,\dots,n\}$ distribution;
    \item 
        $\Delta_{50}$, defined as the value of $50-$th percentile of $\{h_{i,0}:i=1,\dots,n\}$ distribution, i.e. the median;
    \item 
        $\Delta_{75}$, defined as the value of $75-$th percentile of $\{h_{i,0}:i=1,\dots,n\}$ distribution.
\end{enumerate}

We denoted as $A_{\Delta, \mathcal{S}, \mathcal{D}}$ the generic cases, where $\Delta\in\{\Delta_{25}, \Delta_{50}, \Delta_{75}\}$ which represents the threshold for dividing the dataset into train and test, $\mathcal{S}\in \linebreak \{\mathcal{S}_{sup}, \mathcal{S}_{sqrt}\}$ is one of the two modulation functions, $\mathcal{D}\in\{\mathcal{D}_{sup}, \mathcal{D}_{sqrt}\}$ it is one of the nonconformity measures. Combining choice of modulation function, non-conformity measures and the dataset division threshold we have 12 cases study.

\subsection{Simulated data}

\begin{figure}
    \centering
    \includegraphics[width=12.5cm]{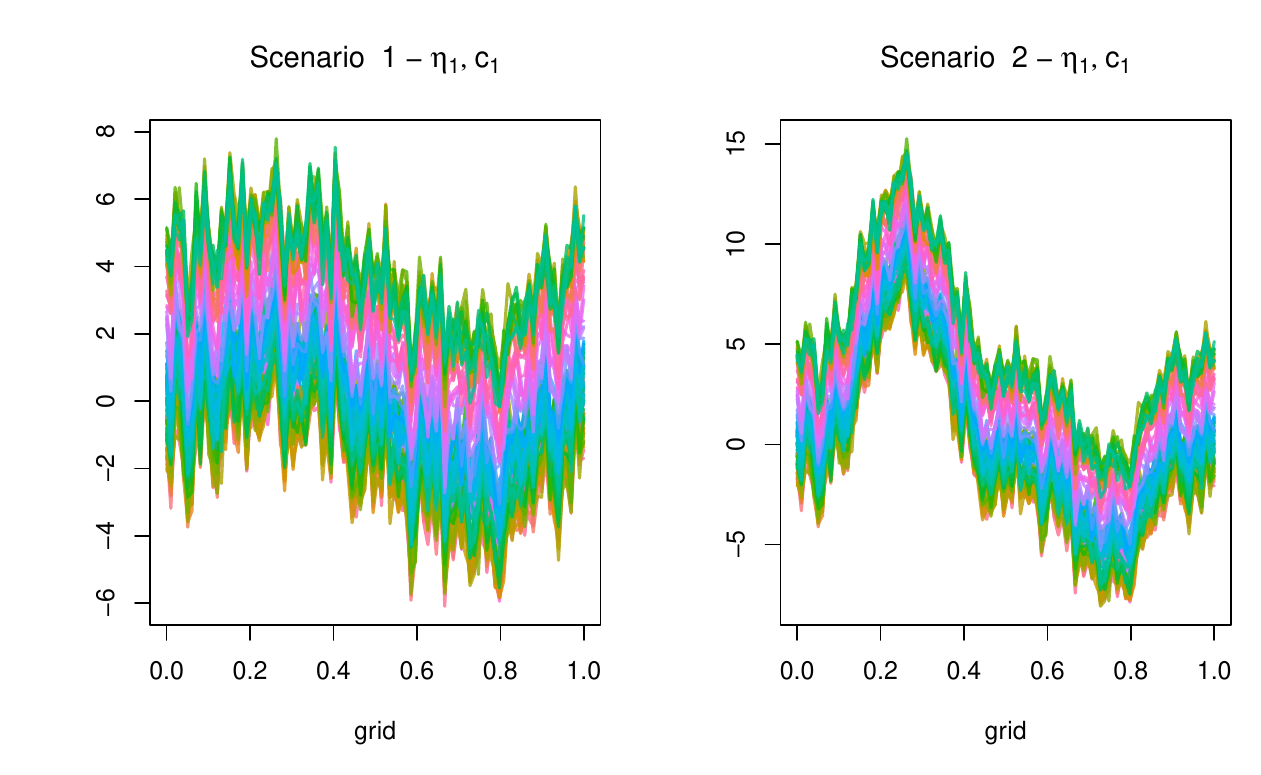}
    \caption{Simulated data with $\eta_1, \: c_1$.}
    \label{fig2}
\end{figure}

We tested our algorithm by creating a simulation framework, as shown in Figure \ref{fig2}, to evaluate the predictive capabilities of the model in various scenarios and datasets. In this study, we generated data using cubic B-splines on a regular spatial grid within the domain $\Omega = [-1, 1] \times [0, 1]$ and the time domain $T = [0, 1]$. We used a B-spline basis on T with $30$ basis functions.

The scenarios are:

\begin{enumerate}
    \item $X_{s}(t) = \mu_{s}(t) + \epsilon_{s}(t)$ (scenario 1);
    \item $X_{s}(t) = [\mu_{s}(t)]^3 + \epsilon_{s}(t)$ (scenario 2).
\end{enumerate}
 
In particular, $\mu_{s}(t)$ is define as follows:
\[\mu_{s}(t) = \frac{1}{2}t + \sin{(2 \pi t)} - 2\sin{(2 \pi t-1)} \log{(2 \pi t + \frac{1}{2})}\]

and $\epsilon_{s}(t)$ is a centered gaussian process with covariance function

\begin{equation} \label{eq:cov}
    C(h)= (1- \eta_i) e^{(-c_j \cdot h) } + \eta_i, \; \;  \; \; i,j = 1,2.
\end{equation}

The formula in \ref{eq:cov} evaluate the covariance between all pairs of points in space, with an exponential decay as distance increases and a nugget term \( \eta \) to account for unexplained variance. By decreasing \( \eta \), small spatial variations are modeled more accurately. The parameter \( c \) represents the decay rate, while \( h \) denotes the distance between points. In our simulations, we've set \( \eta_1 = 0.1,  \; \eta_2 = 0.9 \)  and \( c_1 = 0.1, \;  c_2 = 0.9 \).
For each simulation scenario, we generated functional data with sizes $n = 100$ and Algorithm \ref{algo:1} was applied to each dataset, resulting in twelve different cases.

Table \ref{tab:02}  reports the average of the metrics considered ($Cov{\alpha}_{L}$, $Cov{\alpha}_{G}$, Width, $S_{\alpha}$, Total Time, Mono Time) for Scenario 1, while Table \ref{tab:03} reports the average of the metrics considered for Scenario 2.

In both scenarios 1 and 2, the optimal configuration for achieving the best local and global coverage involves setting the train set size to $\Delta = \Delta_{75}$, using the modulation function defined in equation (\ref{mod_sqrt}), and employing the non-conformance measure from equation (\ref{nnc_sup}). Additionally, for Scenario 1, the most effective parameter settings for defining the covariance function are $\eta = 0.9$ and $ c = 0.9$, while for Scenario 2, the recommended values are $\eta = 0.1$ and $c = 0.9$.

In Scenario 1, when using this parameter setting, the local coverage is $94.82\%$ and the global coverage is $47\%$. In Scenario 2, the local coverage is $92.31\%$ and the global coverage is $57\%$. Regardless of the choice of the size of $\Delta$ from Table \ref{tab:02} and Table \ref{tab:03}, it is evident that the best configuration in terms of local coverage involves setting the modulation function as defined in equation (\ref{mod_sqrt}), and the non-conformity measure as defined in equation (\ref{nnc_sup}). This suggests that it is highly effective to consider the local variability around the prediction $X_{s_0}^*(t)$ and to locally measure the difference between the predictions using the test set and the train set.

\begin{table}[H]
    \centering
    \caption{The average of index integrated $Cov{\alpha}_{L}$, $Cov{\alpha}_{G}$, Width, $S_{\alpha}$, Total Time (TT), Mono Time (MT), evaluated for Simulates datas (Scenario 1), for the all cases of Alg. \ref{algo:1}.}\label{tab:02}
    \begin{tabular}{lcccccccccc}
        \Xhline{2pt}
        {$\Delta$} & {$S$} & {$D$} & {$\eta$} & {$c$} & {Cov$_{\alpha_{L}}$\%} &  {Cov$_{\alpha_G}$ \%} & {Width} &  {$S_{\alpha}$} & {TT} & {MT}  \\
        \Xhline{2pt}
        \multirow{16}{*}{25}  & \multirow{8}{*}{$S_{\text{sup}}$} & \multirow{4}{*}{$D_{\text{sup}}$} & \multirow{2}{*}{$\eta_1$} & $c_1$ & 70.81 & 28 & 15.17 & 39.40 & 474.49 & 4.74 \\
                             &  &  &  & $c_2$ & 79.12 & 28 & 38.74 & 63.18 & 486.10 & 4.86 \\
                             \cline{4-11}
                             &  &  & \multirow{2}{*}{$\eta_2$} & $c_1$ & 72.27 & 23 & 18.03 & 43.00 & 529.30 & 5.29\\
                             &  &  &  & $c_2$ & 78.96 & 26 & 39.33 & 60.80 & 447.47 & 4.47\\
                             
                             \cline{3-11}
                             &  & \multirow{4}{*}{$D_{\text{sqrt}}$} & \multirow{2}{*}{$\eta_1$} & $c_1$ & 60.66 & 26 & 10.54 & 39.85 & 559.31 & 5.59 \\
                             &  &  &  & $c_2$ & 77.07 & 35 & 42.50 & 68.16 & 1126.18 & 11.26 \\
                             \cline{4-11}
                             &  &  & \multirow{2}{*}{$\eta_2$} & $c_1$ & 63.33 & 21 & 13.98 & 39.89 & 1162.63 & 11.63\\
                             &  &  &  & $c_2$& 77.07 & 33 & 42.77 & 67.67 & 1219.38 & 12.19 \\
                   
                             \cline{2-11}
                             & \multirow{8}{*}{$S_{\text{sqrt}}$} & \multirow{4}{*}{$D_{\text{sup}}$} & \multirow{2}{*}{$\eta_1$} & $c_1$ & 73.92 & 29 & 20.25 & 38.34 & 1336.75 & 13.37 \\
                             &  &  &  & $c_2$ & 80.39 & 39 & 48.97 & 71.84 & 678.00 & 6.78 \\
                             \cline{4-11}
                             &  &  & \multirow{2}{*}{$\eta_2$} & $c_1$ & 75.27 & 24 & 25.06 & 41.67 & 1213.21 & 12.13\\
                             &  &  &  & $c_2$ & 80.02 & 37 & 49.55 & 72.36 & 656.18 & 6.56 \\
                             
                             \cline{3-11}
                             &  & \multirow{4}{*}{$D_{\text{sqrt}}$} & \multirow{2}{*}{$\eta_1$} & $c_1$ & 48.17 & 10 & 5.54 & 27.18 & 688.66 & 6.89\\
                             &  &  &  & $c_2$ & 66.48 & 12 & 23.86 & 47.66 & 648.43 & 6.48 \\
                             \cline{4-11}
                             &  &  & \multirow{2}{*}{$\eta_2$} & $c_1$ & 49.44 & 7 & 7.58 & 27.85 & 662.55 & 6.63 \\
                             &  &  &  & $c_2$ & 65.91 & 9 & 23.89 & 47.58 & 651.57 & 6.52 \\
                             \hline
        \Xhline{1.5pt}
        \multirow{16}{*}{50}  & \multirow{8}{*}{$S_{\text{sup}}$} & \multirow{4}{*}{$D_{\text{sup}}$} & \multirow{2}{*}{$\eta_1$} & $c_1$ & 76.35 & 1 & 18.95 & 20.67 & 661.34 & 6.613 \\
                             &  &  &  & $c_2$ & 85.91 & 14 & 48.56 & 53.56 & 657.27 & 6.57 \\
                             \cline{4-11}
                             &  &  & \multirow{2}{*}{$\eta_2$} & $c_1$ & 77.29 & 0 & 22.47 & 24.65 & 650.78 & 6.51 \\
                             &  &  &  & $c_2$ & 85.96 & 11 & 48.36 & 53.21 & 658.78 & 6.59\\
                             
                             \cline{3-11}
                             &  & \multirow{4}{*}{$D_{\text{sqrt}}$} & \multirow{2}{*}{$\eta_1$} & $c_1$ & 64.80 & 2 & 13.19 & 20.58 & 647.89 & 6.48 \\
                             &  &  &  & $c_2$ & 88.89 & 31 & 55.92 & 62.10 & 654.59 & 6.55 \\
                             \cline{4-11}
                             &  &  & \multirow{2}{*}{$\eta_2$} & $c_1$ & 68.78 & 2 & 17.14 & 23.00 & 653.01 & 6.53 \\
                             &  &  &  & $c_2$ & 88.77 & 27 & 55.23 & 62.16 & 655.37 & 6.55 \\
                             
                             \cline{2-11}
                             & \multirow{8}{*}{$S_{\text{sqrt}}$} & \multirow{4}{*}{$D_{\text{sup}}$} & \multirow{2}{*}{$\eta_1$} & $c_1$  & 83.32 & 1 & 24.04 & 25.29 & 651.66 & 6.52 \\
                             &  &  &  & $c_2$ & 90.82 & 32 & 62.54 & 66.13 & 651.98 & 6.52 \\
                             \cline{4-11}
                             &  &  & \multirow{2}{*}{$\eta_2$} & $c_1$ & 84.55 & 1 & 27.78 & 29.56 & 651.18 & 6.51 \\
                             &  &  &  & $c_2$ & 90.94 & 30 & 62.25 & 66.12 & 649.81 & 6.50\\
                             
                             \cline{3-11}
                             &  & \multirow{4}{*}{$D_{\text{sqrt}}$} & \multirow{2}{*}{$\eta_1$} & $c_1$ & 48.13 & 1 & 8.28 & 25.82 & 651.02 & 6.51\\
                             &  &  &  & $c_2$ & 77.61 & 4 & 36.08 & 53.98 & 677.80 & 6.78 \\
                             \cline{4-11}
                             &  &  & \multirow{2}{*}{$\eta_2$} & $c_1$ & 51.17 & 0 & 10.81 & 27.74 & 651.29 & 6.51 \\
                             &  &  &  & $c_2$& 77.18 & 1 & 36.02 & 52.34 & 655.08 & 6.55 \\
                             \hline
        \Xhline{1.5pt}
        \multirow{16}{*}{75}  & \multirow{8}{*}{$S_{\text{sup}}$} & \multirow{4}{*}{$D_{\text{sup}}$} & \multirow{2}{*}{$\eta_1$} & $c_1$ & 79.57 & 1 & 20.49 & 22.23 & 578.14 & 5.78 \\
                             &  &  &  & $c_2$ & 88.90 & 7 & 57.13 & 62.38 & 504.94 & 5.05 \\
                             \cline{4-11}
                             &  &  & \multirow{2}{*}{$\eta_2$} & $c_1$ & 82.87 & 2 & 24.55 & 26.26 & 490.86 & 4.91\\
                             &  &  &  & $c_2$& 90.74 & 5 & 57.43 & 62.64 & 521.15 & 5.21 \\
                             
                             \cline{3-11}
                             &  & \multirow{4}{*}{$D_{\text{sqrt}}$} & \multirow{2}{*}{$\eta_1$} & $c_1$ & 70.38 & 1 & 16.06 & 19.69 & 475.07 & 4.75 \\
                             &  &  &  & $c_2$ & 93.39 & 43 & 76.22 & 84.98 & 503.24 & 5.03 \\
                             \cline{4-11}
                             &  &  & \multirow{2}{*}{$\eta_2$} & $c_1$ & 76.92 & 3 & 21.03 & 24.24 & 480.38 & 4.80 \\
                             &  &  &  & $c_2$ & 94.34 & 47 & 76.63 & 82.78 & 505.69 & 5.06 \\
                           
                             \cline{2-11}
                             & \multirow{8}{*}{$S_{\text{sqrt}}$} & \multirow{4}{*}{$D_{\text{sup}}$} & \multirow{2}{*}{$\eta_1$} & $c_1$ & 88.47 & 1 & 27.96 & 29.65 & 473.03 & 4.73 \\
                             &  &  &  & $c_2$ & 93.38 & 47 & 77.50 & 83.30 & 534.46 & 5.34 \\
                             \cline{4-11}
                             &  &  & \multirow{2}{*}{$\eta_2$} & $c_1$ & 90.90 & 5 & 33.64 & 34.55 & 485.10 & 4.85 \\
                             &  &  &  & $c_2$ & \textbf{94.82} & \textbf{47} & \textbf{78.12} & \textbf{83.21} & \textbf{621.14} & \textbf{6.21} \\
                             
                             \cline{3-11}
                             &  & \multirow{4}{*}{$D_{\text{sqrt}}$} & \multirow{2}{*}{$\eta_1$} & $c_1$ & 53.94 & 1 & 10.92 & 21.52 & 648.92 & 6.49 \\
                             &  &  &  & $c_2$ & 86.21 & 7 & 52.61 & 59.93 & 619.43 & 6.19 \\
                             \cline{4-11}
                             &  &  & \multirow{2}{*}{$\eta_2$} & $c_1$ & 61.06 & 2 & 14.35 & 23.74 & 657.73 & 6.58 \\
                             &  &  &  & $c_2$ & 87.95  & 10 & 52.67 & 60.00 & 451.27 & 4.51 \\
                             \hline
    \end{tabular}
\end{table}

\begin{table}[H]
    \centering
    \caption{The average of index integrated $Cov{\alpha}_{L}$, $Cov{\alpha}_{G}$, Width, $S_{\alpha}$, Total Time (TT), Mono Time (MT), evaluated for Simulates datas (Scenario 2), for the all cases of Alg. \ref{algo:1}.}\label{tab:03}
    \begin{tabular}{lcccccccccc}
        \Xhline{2pt}
        {$\Delta$} & {$S$} & {$D$} & {$\eta$} & {$c$} & {Cov${\alpha}_{L}$\%} &  {Cov${\alpha}_G$ \%} & {Width} &  {$S_{\alpha}$} & {TT} & {MT}  \\
        \Xhline{2pt}
        \multirow{16}{*}{25}  & \multirow{8}{*}{$S_{\text{sup}}$} & \multirow{4}{*}{$D_{\text{sup}}$} & \multirow{2}{*}{$\eta_1$} & $c_1$ &  66.80 & 28 & 13.57 & 37.60 & 482.49 & 4.82 \\
                             &  &  &  & $c_2$ & 72.62 & 29 & 34.48 & 64.96 & 448.80 & 4.49\\
                             \cline{4-11}
                             &  &  & \multirow{2}{*}{$\eta_2$} & $c_1$ & 67.66 & 24 & 15.91 & 38.39 & 475.26 & 4.75\\
                             &  &  &  & $c_2$ & 73.29 & 29 & 35.47 & 63.53 & 447.25 & 4.47\\
                             
                             \cline{3-11}
                             &  & \multirow{4}{*}{$D_{\text{sqrt}}$} & \multirow{2}{*}{$\eta_1$} & $c_1$ & 56.25 & 24 & 8.33 & 38.37 & 477.54 & 4.7 \\
                             &  &  &  & $c_2$ & 68.76 & 31 & 36.52 & 73.36 & 447.80 & 4.48\\
                             \cline{4-11}
                             &  &  & \multirow{2}{*}{$\eta_2$} & $c_1$ & 58.05 & 21 & 10.91 & 39.00 & 477.43 & 4.77\\
                             &  &  &  & $c_2$ & 69.12 & 31 & 35.84 & 73.01 & 449.52 & 4.5 \\
                   
                             \cline{2-11}
                             & \multirow{8}{*}{$S_{\text{sqrt}}$} & \multirow{4}{*}{$D_{\text{sup}}$} & \multirow{2}{*}{$\eta_1$} & $c_1$ & 69.01 & 24 & 16.81 & 37.93 & 480.08 & 4.80 \\
                             &  &  &  & $c_2$ & 73.97 & 34 & 43.13 & 68.33 & 447.25 & 4.47 \\
                             \cline{4-11}
                             &  &  & \multirow{2}{*}{$\eta_2$} & $c_1$ & 69.67 & 26 & 19.86 & 41.06 & 478.82 & 4.79\\
                             &  &  &  & $c_2$ & 74.37 & 33 & 43.28 & 69.97 & 450.99 & 4.51 \\
                             
                             \cline{3-11}
                             &  & \multirow{4}{*}{$D_{\text{sqrt}}$} & \multirow{2}{*}{$\eta_1$} & $c_1$ & 45.82 & 9 & 4.98 & 30.17 & 476.74 & 4.77\\
                             &  &  &  & $c_2$ & 58.98 & 9 & 22.38 & 49.95 & 449.26 & 4.49\\
                             \cline{4-11}
                             &  &  & \multirow{2}{*}{$\eta_2$} & $c_1$ & 46.40 & 6 & 6.50 & 30.54 & 478.96 & 4.79 \\
                             &  &  &  & $c_2$ & 59.45 & 7 & 22.31 & 50.43 & 447.90 & 4.48 \\
                             \hline
        \Xhline{1.5pt}
        \multirow{16}{*}{50}  & \multirow{8}{*}{$S_{\text{sup}}$} & \multirow{4}{*}{$D_{\text{sup}}$} & \multirow{2}{*}{$\eta_1$} & $c_1$ & 71.43 & 1 & 17.90 & 19.28 & 446.33 & 4.46\\\
                             &  &  &  & $c_2$ & 84.43 & 12 & 46.19 & 52.67 & 428.18 & 4.28 \\
                             \cline{4-11}
                             &  &  & \multirow{2}{*}{$\eta_2$} & $c_1$ & 74.16 & 3 & 21.35 & 22.97 & 450.64 & 4.51 \\
                             &  &  &  & $c_2$ & 83.02 & 11 & 46.39 & 52.91 & 427.96 & 4.28\\
                             
                             \cline{3-11}
                             &  & \multirow{4}{*}{$D_{\text{sqrt}}$} & \multirow{2}{*}{$\eta_1$} & $c_1$ & 59.53 & 1 & 12.61 & 20.34 & 445.99 & 4.46 \\
                             &  &  &  & $c_2$ & 85.88 & 29 & 55.29 & 63.44 & 428.71 & 4.29 \\
                             \cline{4-11}
                             &  &  & \multirow{2}{*}{$\eta_2$} & $c_1$ & 65.13 & 3 & 16.01 & 23.17 & 449.66 & 4.50 \\
                             &  &  &  & $c_2$ & 84.99 & 24 & 54.91 & 64.12 & 426.69 & 4.27 \\
                             
                             \cline{2-11}
                             & \multirow{8}{*}{$S_{\text{sqrt}}$} & \multirow{4}{*}{$D_{\text{sup}}$} & \multirow{2}{*}{$\eta_1$} & $c_1$  & 80.49 & 0 & 23.02 & 24.64 & 449.41 & 4.49 \\
                             &  &  &  & $c_2$ & 89.56 & 29 & 59.23 & 63.98 & 423.23 & 4.23 \\
                             \cline{4-11}
                             &  &  & \multirow{2}{*}{$\eta_2$} & $c_1$ & 82.52 & 3 & 27.11 & 28.56 & 446.53 & 4.47 \\
                             &  &  &  & $c_2$ & 88.98 & 29 & 58.97 & 64.18 & 436.40 & 4.36\\
                             
                             \cline{3-11}
                             &  & \multirow{4}{*}{$D_{\text{sqrt}}$} & \multirow{2}{*}{$\eta_1$} & $c_1$ & 43.51 & 0 & 7.93 & 25.21 & 446.84 & 4.47 \\
                             &  &  &  & $c_2$ & 75.82 & 6 & 35.79 & 51.24 & 445.12 & 4.45  \\
                             \cline{4-11}
                             &  &  & \multirow{2}{*}{$\eta_2$} & $c_1$ & 49.11 & 3 & 10.09 & 26.67 & 455.21 & 4.55 \\
                             &  &  &  & $c_2$ & 75.08 & 5 & 35.93 & 52.01 & 427.00 & 4.27 \\
                             \hline
        \Xhline{1.5pt}
        \multirow{16}{*}{75}  & \multirow{8}{*}{$S_{\text{sup}}$} & \multirow{4}{*}{$D_{\text{sup}}$} & \multirow{2}{*}{$\eta_1$} & $c_1$ & 69.82 & 0 & 18.78 & 20.77 & 420.87 & 4.21 \\
                             &  &  &  & $c_2$ & 88.54 & 13 & 55.92 & 59.62 & 419.86 & 4.20 \\
                             \cline{4-11}
                             &  &  & \multirow{2}{*}{$\eta_2$} & $c_1$ & 79.49 & 2 & 22.96 & 24.47 & 422.76 & 4.23\\
                             &  &  &  & $c_2$ & 88.54 & 13 & 55.92 & 59.62 & 419.86 & 4.20 \\
                             
                             \cline{3-11}
                             &  & \multirow{4}{*}{$D_{\text{sqrt}}$} & \multirow{2}{*}{$\eta_1$} & $c_1$ & 88.47 & 9 & 56.21 & 59.49 & 419.27 & 4.19 \\
                             &  &  &  & $c_2$ & 70.86 & 3 & 18.70 & 22.97 & 422.01 & 4.22 \\
                             \cline{4-11}
                             &  &  & \multirow{2}{*}{$\eta_2$} & $c_1$ & 59.62 & 1 & 13.98 & 19.67 & 417.16 & 4.17 \\
                             &  &  &  & $c_2$ & 91.59 & 52 & 73.27 & 78.22 & 422.07 & 4.22 \\
                           
                             \cline{2-11}
                             & \multirow{8}{*}{$S_{\text{sqrt}}$} & \multirow{4}{*}{$D_{\text{sup}}$} & \multirow{2}{*}{$\eta_1$} & $c_1$ & 77.45 & 0 & 25.45 & 28.45 & 417.13 & 4.17\\
                             &  &  &  & $c_2$ & \textbf{92.31} & \textbf{56} & \textbf{75.56} & \textbf{80.82} & \textbf{418.29} & \textbf{4.18}\\
                             \cline{4-11}
                             &  &  & \multirow{2}{*}{$\eta_2$} & $c_1$ & 85.39 & 6 & 30.97 & 33.55 & 421.98 & 4.22 \\
                             &  &  &  & $c_2$ & 92.06 & 57 & 75.13 & 80.63 & 419.65 & 4.20 \\
                             
                             \cline{3-11}
                             &  & \multirow{4}{*}{$D_{\text{sqrt}}$} & \multirow{2}{*}{$\eta_1$} & $c_1$ & 45.08 & 1 & 9.18 & 21.20 & 413.91 & 4.14 \\
                             &  &  &  & $c_2$ & 86.75 & 9 & 49.83 & 55.59 & 416.12 & 4.16 \\
                             \cline{4-11}
                             &  &  & \multirow{2}{*}{$\eta_2$} & $c_1$ & 55.87 & 2 & 12.29 & 20.69 & 419.80 & 4.20 \\
                             &  &  &  & $c_2$ & 86.42 & 10 & 49.58 & 55.90 & 413.43 & 4.13 \\
                             \hline
    \end{tabular}
\end{table}


\section{Real case study}
\label{sec:5}

In order to evaluate the effectiveness of the new method, we compare the results obtained with a new approach to the classical bootstrap methods proposed by \cite{Ramsay:FDA}. We  perform tests using actual data that is widely recognized in the academic community, such as the Canadian temperature dataset \cite{Franco-Villoria:17,Giraldo:11,Ramsay:FDA}. This particular dataset comprises daily annual mean temperature measurements from 35 meteorological stations located in Canada's Maritimes Provinces. Our approach involved using the method to forecast temperatures at each of the $35$ Canadian meteorological stations. This process required executing the procedure $35$ times, with each iteration excluding the station for which the prediction was being made. To model the data, we employed a Fourier basis with $65$ bases, as the temperature data exhibit distinct cyclical patterns.

\begin{table}[ht]
    \centering
    \caption{The average of index integrated $Cov{\alpha}_{L}$, $Cov{\alpha}_{G}$, Width, $S_{\alpha}$, Total Time (TT), Mono Time (MT), evaluated for the $35$ meteorological stations, for the 12 cases of Alg. \ref{algo:1}.}\label{tab:01}
    \begin{tabular}{lcccccccccc}
        \Xhline{2pt}
        {$\Delta$} & {$S$} & {$D$} & & {Cov${\alpha}_{L}$\%} &  {Cov${\alpha}_G$ \%} & {Width} & & {$S_{\alpha}$} & {TT} & {MT}  \\
        \Xhline{2pt}
        \multirow{4}{*}{25}  & \multirow{2}{*}{$S_{\text{sup}}$} & $D_{\text{sup}}$ & & 81.33  & 8.57 & 1257.53 & & 1953.54 & 19.75 & 0.56 \\
                             &  & $D_{\text{sqrt}}$ & & 76.36  & 25.71 & 987.73 & & 2325.98 & 20.00 & 0.57 \\
                              \cline{2-11}
                             & \multirow{2}{*}{$S_{\text{sqrt}}$} & $D_{\text{sup}}$ & & 85.12 &  17.14 & 1483.05 & & 2208.64 & 21.03 & 0.60 \\
                             &  & $D_{\text{sqrt}}$ && 60.82 & 20 & 553.60 & & 3149.88 & 19.90 & 0.56 \\
        \Xhline{1.5pt}
        \multirow{4}{*}{50}  & \multirow{2}{*}{$S_{\text{sup}}$} & $D_{\text{sup}}$ & & 83.95 &  17.14 & 1880.69 & & 2259.07 & 18.58 & 0.53 \\
                             &  & $D_{\text{sqrt}}$ & & 84.80 &  25.71 & 2447.85 & & 2519.67 & 18.79 & 0.53 \\
                              \cline{2-11}
                             & \multirow{2}{*}{$S_{\text{sqrt}}$} & $D_{\text{sup}}$ & & \textbf{90.49} &  \textbf{22.85} & \textbf{2465.30} & & \textbf{2500.36} & \textbf{18.83} & \textbf{0.53} \\
                             &  & $D_{\text{sqrt}}$ &  & 75.49 &  14.28 & 1319.61 & & 2613.43 & 18.67 & 0.53\\
        \Xhline{1.5pt}
        \multirow{4}{*}{75}  & \multirow{2}{*}{$S_{\text{sup}}$} & $D_{\text{sup}}$ & & 81.62 &  11.42 & 2445.81 & & 3080.48 & 12.91 & 0.36 \\
                             &  & $D_{\text{sqrt}}$ && 86.71 &  42.85 & 3847.07 & & 4637.49 & 12.82& 0.36 \\
                             \cline{2-11}
                             & \multirow{2}{*}{$S_{\text{sqrt}}$} & $D_{\text{sup}}$ & & 88.98 &  20 & 2960.99 & & 3389.05 & 13.05  & 0.37 \\
                             &  & $D_{\text{sqrt}}$ & & 80.90 &  25.71 & 2362.84 & & 3834.42 & 12.99 & 0.37 \\
        \hline
        Bootstrap & & &  & 96.27 & 71.43 & 5657.023 & & 5753.30 & 998.16 & 28.52\\
        \Xhline{2pt}
    \end{tabular}
    \label{tab:gof}
\end{table}

Table \ref{tab:01} summarises the main results obtained from the prediction of 35 meteorological stations using the 12 cases of our procedure .
It is obvious that the cases $A_{\Delta_{25}, \mathcal{S}_{sqrt}, \mathcal{D}_{sup}}$, $A_{\Delta_{50}, \mathcal{S}_{sqrt}, \mathcal{D}_{sup}}$, and $A_{\Delta_{75}, \mathcal{S}_{sqrt}, \mathcal{D}_{sup}}$ exhibit the best performance in terms of both $Cov\alpha_{L}$ and $Cov\alpha_G$. This highlights the importance of selecting an appropriate combination of non-conformity measure and modulation function to enhance the performance of Algorithm \ref{algo:1}. In general, the best performances are obtained by choosing proximity thresholds $\Delta_{50}$. In the last row of the table were reported the results of the same indices using boostrap method. Despite higher global and local coverage in bootstrap methods, computational time for prediction and band construction is superior in conformal prediction. In conformal prediction, the band width is narrower compared to bootstrap methods. The bootstrap method is more computationally costly because it involves repeated resampling of the dataset, whereas the Conformal Prediction method only requires dividing the dataset.

Furthermore, in the case of the bootstrap method we have to make assumptions about the process: we have to require that our data are independent and identically distributed, while in the case of conformal prediction the only assumption we make is that the data are spatially exchangeable. Another advantage of the Conformal Prediction method is that it introduces variability by adding a test set element to the training set. In contrast, the bootstrap method simplifies the data set. In addition, while the bootstrap method creates the band by piecing together confidence intervals built around individual time points of the curve, our procedure constructs the band directly around the curve.

\begin{figure}[ht]
    \centering
    \includegraphics[width=0.5\linewidth]{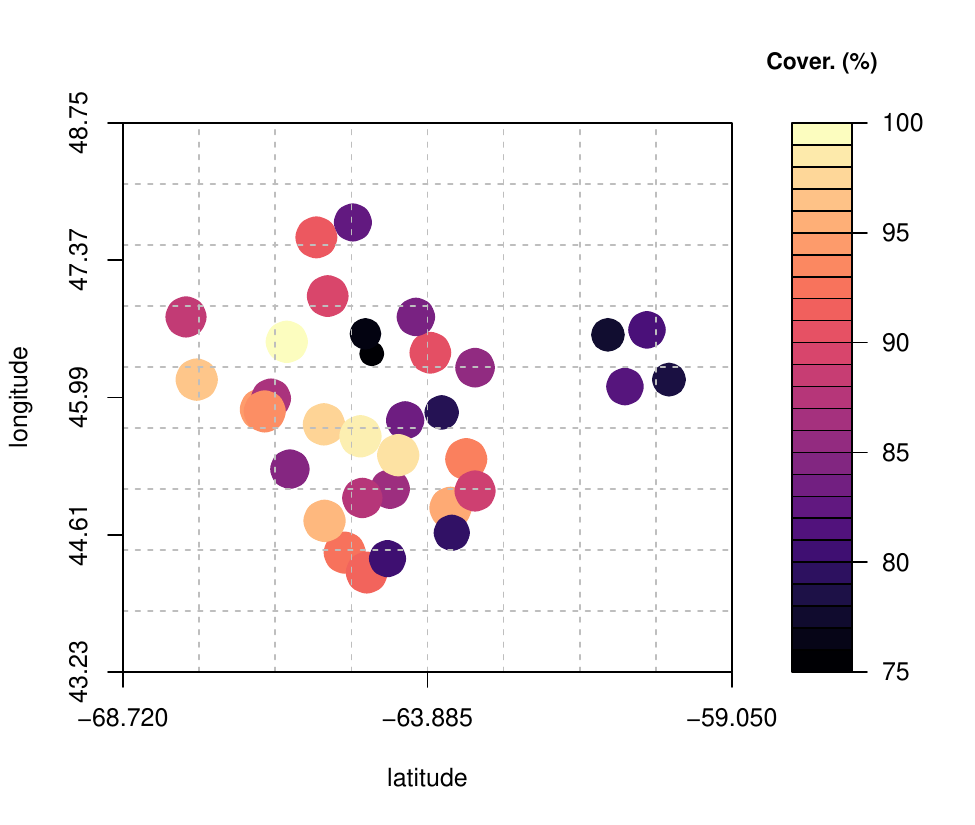}
    \caption{The 35 meteorological stations in Canada’s Maritimes Provinces. The filler and the size of points on the map are releted to Cov$_{\alpha_{L}}$ of $A_{\Delta_{50}, \mathcal{S}_{sqrt}, \mathcal{D}_{sup}}$.}
    \label{fig:00}
\end{figure}

Figure \ref{fig:00} represents a map of 35 weather stations in the Maritime Provinces of Canada. Each station is represented by a point, with varying sizes and fill colours that correspond to the calculated Cov${\alpha_{L}}$ for $A_{\Delta_{50}, \mathcal{S}_{sqrt}, \mathcal{D}_{sup}}$. Based on the map, it is evident that the majority of the points indicate a high local coverage of around $90\%$.

\begin{figure}[H]
    \centering
    \begin{subfigure}[b]{0.45\textwidth}
        \includegraphics[width=\textwidth]{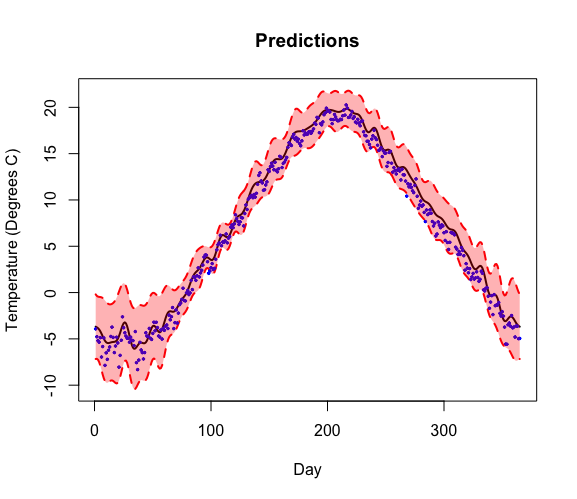}
        \caption{}
        \label{fig:sub1}
    \end{subfigure}
    \quad
    \begin{subfigure}[b]{0.45\textwidth}
        \includegraphics[width=\textwidth]{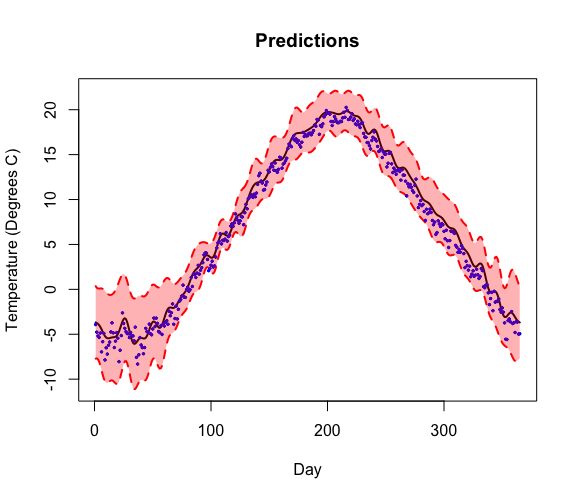}
        \caption{}
        \label{fig:sub2}
    \end{subfigure}
    \\
    \begin{subfigure}[b]{0.45\textwidth}
        \includegraphics[width=\textwidth]{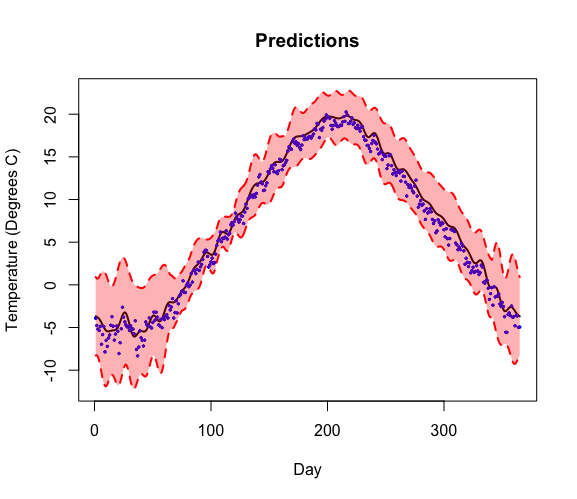}
        \caption{}
        \label{fig:sub3}
    \end{subfigure}
    \quad
    \begin{subfigure}[b]{0.45\textwidth}
        \includegraphics[width=\textwidth]{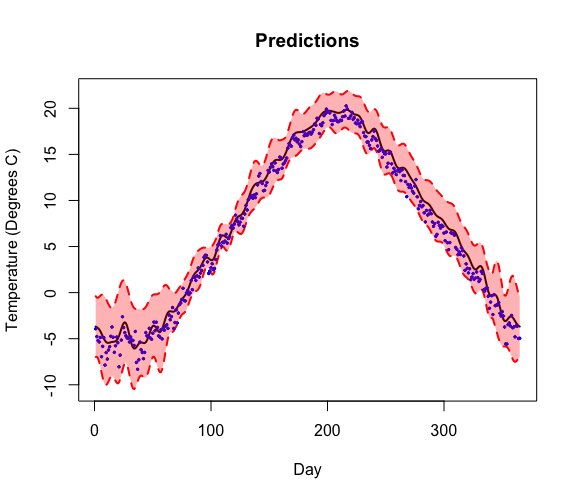}
        \caption{}
        \label{fig:sub4}
    \end{subfigure}
    \caption{Forecast of a single weather station: a)$A_{\Delta_{50}, \mathcal{S}_{sqrt}, \mathcal{D}_{sqrt}}$, b)$A_{\Delta_{50}, \mathcal{S}_{sqrt}, \mathcal{D}_{sup}}$ c)$A_{\Delta_{50}, \mathcal{S}_{sup}, \mathcal{D}_{sqrt}}$ and d)$A_{\Delta_{50}, \mathcal{S}_{sup}, \mathcal{D}_{sup}}$. }
    \label{fig1}
\end{figure}

Figure \ref{fig1} displays the prediction bands for a single meteorological station along with its corresponding prediction with $\Delta_{50}$: the red curves represent the prediction bands, the black curve represents the prediction of the meteorological station under consideration, and the blue dots denote the actual temperature data from the Canadian station being analysed. As we can see in the Figure \ref{fig:sub2}, the prediction bands obtained with $A_{\Delta_{50}, \mathcal{S}_{sqrt}, \mathcal{D}_{sup}}$  successfully encompass all the actual data points, unlike $A_{\Delta_{50}, \mathcal{S}_{sqrt}, \mathcal{D}_{sqrt}}$ in Figure \ref{fig:sub1}, $A_{\Delta_{50}, \mathcal{S}_{sup}, \mathcal{D}_{sqrt}}$ in Figure \ref{fig:sub3} and $A_{\Delta_{50}, \mathcal{S}_{sup}, \mathcal{D}_{sup}}$ in Figure \ref{fig:sub4}.

\section{Conclusion}
        \label{sec:6}

In this study, we addressed the problem of evaluating uncertainty in the prediction of spatial functional data using Kriging methods for prediction and Conformal Prediction to assess uncertainty. By proposing a new distribution-free approach based on Conformal Prediction, we introduced a flexible and computationally efficient alternative to traditional methods such as the bootstrap. Our method was tested on both real and simulated datasets, demonstrating clear advantages in terms of computational cost, required assumptions, and prediction accuracy. 
In contrast to the bootstrap method, which is computationally expensive and assumes strict distributional properties such as independence and identical distribution (i.i.d.), the Conformal Prediction approach requires only spatial interchangeability. This assumption allows our method to be more widely applicable across a wider range of datasets. Constructing prediction bands directly around the entire curve, rather than around individual points, provides a more holistic and robust uncertainty estimate.

The practical benefits of our method were highlighted through its application to real-world spatial functional data, where it outperformed traditional methods by offering tighter and more reliable uncertainty bounds without compromising prediction accuracy. Moreover, the computational efficiency of Conformal Prediction makes it a particularly attractive option for large-scale applications where repeated resampling is impractical.

Future research could focus on extending this method by exploring different non-conformity measures and modulation functions to further enhance the flexibility and generalizability of Conformal Prediction in the context of spatial functional Kriging. Such developments could open new avenues for improving prediction accuracy and uncertainty quantification in spatial functional data analysis.
\bibliography{bibliograf}
\bibliographystyle{plain}

\end{document}